\documentclass[pra,twocolumn,longbibliography]{revtex4-2}
\usepackage{amsmath}
\usepackage{amssymb}
\usepackage{graphicx} 

\usepackage[usenames, dvipsnames]{color}
\usepackage[colorlinks=true,citecolor=blue,urlcolor  = purple]{hyperref}

\newcommand{\mean}[1]{\langle #1 \rangle}

\newcommand{\LSQ}{\mbox{\scriptsize LSQ}}
\newcommand{\RE}{\mbox{\scriptsize LRE}}
\newcommand{\IBU}{\mbox{\scriptsize IBU}}
\usepackage{xcolor}
\newcommand{\red}[1]{{\color{black} #1}}

\begin{document}

\title{Information theoretic approach to readout error mitigation for quantum computers}
\author{Hai-Chau Nguyen}
\affiliation{Naturwissenschaftlich-Technische Fakultät, 
Universität Siegen, Walter-Flex-Straße 3, 57068 Siegen, Germany}
\email{chau.nguyen@uni-siegen.de}
\date{\today}

\begin{abstract}
    We show that the method of iterative bayesian unfolding for mitigating readout errors in quantum computers can be derived from an information theoretic analysis. 
    This inspires more flexible applications of this error mitigation scheme.
    In particular, we distinguish between structural mitigation and unstructural mitigation. 
    Structural mitigation addresses nearly deterministic quantum computation, where the computer is expected to output a single or few outcome bitstrings. It is shown that the readout errors alone can be corrected by few repetitions of the computation.
    In contrast, unstructural mitigation is designed for quantum simulation, where the computer outputs bitstrings broadly distributed. 
    In this case, one is interested in mitigating certain observables of interest. 
    As most observables of interest are dependent on few bits and not the whole bitstring, it is sufficient to mitigate the marginal distributions over these dependent bits. 
    As long as the cross-talk of readout errors can be ignored, it is shown that the iterative bayesian unfolding applied locally for these marginal distributions gives similar results as mitigation using least squared errors.  
    We illustrate our analysis using the data of the preparation of the GHZ state in a 127-qubit quantum computer.
\end{abstract}

\maketitle

\section{Introduction}

The first generation of quantum computers with several to few hundred qubits has now arrived after several decades of intense research effort~\cite{arute_quantum_2019,zhong2021a,wu2021a,ebadi2021a}. 
On the one hand, this represents the first promising step towards the future quantum technology.
On the other hand, these contemporary devices are still relatively small and suffer from serious noise~\cite{nisq_preskill_2018}. 
As a result, the practical advantages of quantum computation are yet to be unambiguously demonstrated. 
As a response, a subfield of quantum information theory has emerged, aiming at mitigating the errors for near-term quantum computers~\cite{Temme2017a,Li2017a,Suguru2018a,Matthew2019a,cai_quantum_2023}, hoping for finding  their practical applications despite their small sizes and noisy operations~\cite{Kandala2019a,Larose2022a}.

Recent effort has pinpointed the classical errors in reading the outcomes of the measurements in the near-term quantum computers as an important part of the errors ~\cite{Chow2010a,Chow2012a,Ryan2015,Yanzhu2019a,mooney_whole-device_2021}. 
Suppose by the end of its operation, the quantum computer performs a measurement on the qubits, which ideally yields an outcome bitstring of $\pmb{\xi}=(\xi_1,\xi_2,\ldots,\xi_n)$.
However, because of errors occurring during the readout process, the registered outcome $\pmb{s}=(s_1,s_2,\ldots,s_n)$ actually differs from this ideal outcome $\pmb{\xi}$. 
In this work, we limit ourselves to the commonly used independent readout noise model~\cite{nation_scalable_2021,srinivasan_scalable_2022}.  
Under the assumption of the independent readout noise model, each of the bit $s_k$ is the result of randomly flipping $\xi_k$ independently from each other with transition rate $R_k(s_k \vert \xi_k)$.
Consequently, the transition matrix of observing a bistring $\pmb{s}$ given an ideal bitstring $\pmb{\xi}$ is given by
\begin{equation}
    R(\pmb{s}|\pmb{\xi}) = \prod_{k=1}^{n} R_k (s_k \vert \xi_k).
    \label{eq:noise}
\end{equation}
During calibration, the flipping rates of single qubits $R_k(s_k \vert \xi_k)$ can be estimated with high accuracy.

Often the quantum computation is repeated $M$ times. 
These repeated runs yield then $M$ observed bitstrings  $\{\pmb{s}_1,\pmb{s}_2,\ldots,\pmb{s}_M\}$. 
The data can be used to estimate the probability distribution for the observed bitstrings $P(\pmb{s})$ as 
\begin{equation}
 P(\pmb{s}) \approx \frac{1}{M} \sum_{\mu=1}^{M} \delta_{\pmb{s},\pmb{s}_\mu}, 
 \label{eq:statistics}
\end{equation}
where $\delta$ denotes the Kronecker symbol.
Given the distribution over the observed bitstrings $P(\pmb{s})$ and the noise model~\eqref{eq:noise}, the challenge is to estimate the actual probability distribution $Q(\pmb{\xi})$ over the hidden ideal outcomes $\pmb{\xi}$. 

Many methods have been devoted to mitigation of these readout errors, see Refs.~\cite{bravyi_mitigating_2021,nachman_unfolding_2020,srinivasan_scalable_2022,nation_scalable_2021,berg_model-free_2022} and the references therein.
Naively, the distribution $Q(\pmb{\xi})$ over the ideal outcome bitstrings  can be obtained by minimising the sum of squared errors between  $P(\pmb{s})$ and   $\sum_{\pmb{\xi}} R(\pmb{s}|\pmb{\xi}) Q(\pmb{\xi})$,
\begin{equation}
Q_{\LSQ} = \arg \min_Q \sum_{\pmb{s}} [ P(\pmb{s}) - \sum_{\pmb{\xi}} R(\pmb{s}|\pmb{\xi}) Q(\pmb{\xi}) ]^2.
\label{eq:sos}
\end{equation}
This amounts to invert the transition matrix $R(\pmb{s}|\pmb{\xi})$ and apply to the vector of observed probabilities $P(\pmb{s})$,
\begin{equation}
    Q_{\LSQ} (\pmb{\xi}) = \sum_{\pmb{s}} R^{-1}(\pmb{\xi} \vert \pmb{s}) P(\pmb{s}),
\end{equation}
where $R^{-1}(\pmb{\xi} \vert \pmb{s})$ denotes the inverse of $R(\pmb{s}|\pmb{\xi})$ as a matrix.

The criticism toward this method of least squared errors is that $Q_{\LSQ} (\pmb{s})$ under low sampling ($M \ll 2^n$) often contains negative values. As such it cannot be regarded as a probability distribution, but rather considered as quasi-probabilities at the best. 
Lately, it has been realised that a method widely used in analysing data of high energy physics experiments known as Iterative Bayesian Unfolding (IBU) can address this problem~\cite{nachman_unfolding_2020}. 
In IBU, the distribution $Q_{\IBU}(\pmb{\xi})$ is found by following an iterative update rule
\begin{equation}
    Q^{(r+1)} (\pmb{\xi}) = \sum_{\pmb{s}} P(\pmb{s}) \frac{ R(\pmb{s}|\pmb{\xi}) Q^{(r)}(\pmb{\xi})}{\sum_{\pmb{\xi}'} R(\pmb{s}|\pmb{\xi}') Q^{(r)}(\pmb{\xi}')}
    \label{eq:ibu}
\end{equation}
until convergence.
It is clear that once initiated with positive values, $Q^{(r)} (\pmb{\xi})$ consistently remains positive.
Interestingly, the authors of Ref.~\cite{srinivasan_scalable_2022} have pointed out that this procedure can be understood as an expression of the general Expectation--Maximisation algorithm known in statistical analysis and machine learning~\cite{Dempster1977a}.

It is observed that both approaches~\eqref{eq:sos} and~\eqref{eq:ibu} attempt to infer all probability weights $Q(\pmb{\xi})$.
One might therefore naturally question their scalability for large system size $n$, since the total number of bitstrings increases exponentially as $2^n$. 
As a possible resolution, it was suggested that $Q(\pmb{\xi})$ is often sparse, and close to the observed data bitstrings~\cite{nation_scalable_2021}. 
Consequently, one can assume that $Q$ is supported only around the observed data bitstrings upto certain Hamming distance~\cite{nation_scalable_2021}.
This approach has been used to adapt the least square error mitigation~\cite{nation_scalable_2021}  and also the IBU mitigation~\cite{srinivasan_scalable_2022} to work with relatively longer outcome bitstrings.

However, for even a larger number of qubits, the sampled observed bitstrings are largely outnumbered by all the possible bitstrings, $M \ll 2^n$.
In this case, even $P(\pmb{s})$ may not be estimated with a reasonable accuracy. 
Thus it is also expected that $Q(\pmb{\xi})$ cannot be inferred to a high accuracy by direct application of~\eqref{eq:sos} or~\eqref{eq:ibu}.
To have a closer look at this problem, we may classify a quantum computation process in two large categories, (nearly) deterministic quantum computation and quantum simulation. 

For deterministic quantum computation,  the quantum computer ideally outputs a single outcome bitstring. 
In this case, if the errors other that the readout errors can be ignored, it is essentially a problem of classical communication of a single bitstring through a noisy communication channel~\cite{Cover1991a}.
One might expect that  methods of classical error correction can be used and several repetitions of the computation should be sufficient to uncover the ideal outcome bitstring.
A bit more generally, the ideal quantum computer can output few specific bitstrings, each with certain probability.
Formally, this means that $Q(\pmb{\xi})$ can be assumed to be supported on few (unknown) bitstrings. 
We show that a slightly generalised version of the IBU mitigation can be used to infer $Q(\pmb{\xi})$ in this case. 
This scenario in fact resembles the concepts of inference of latent variables in machine learning literature~\cite{Bishop2006a}.
We refer to this as \emph{structural mitigation} of the distribution.

The situation is different for quantum simulation. 
In this case the quantum simulator outputs a specific quantum state in every run but subsequently repeated measurements yield outcome bitstrings broadly distributed over all possible bitstrings. 
Fortunately, in this case one is most often not interested the exponential number of probability weights $Q(\pmb{\xi})$ themselves.
Indeed, most often one is interested in inferring certain observables from the distribution $Q(\pmb{\xi})$~\cite{bravyi_mitigating_2021}.
As no structure of the distribution $Q(\pmb{\xi})$ is assumed, we refer to this as \emph{unstructural mitigation}.
In fact, this shift of paradigm from inferring the distribution itself to inferring its associated observables has been also made for quantum states themselves, where shadow tomography is suggested to replace complete state tomography for systems of large number of qubits~\cite{original_shadow_huang_2020}.
It has been soon realised that the least square approach~\eqref{eq:sos} can be adapted to inferring many observables efficiently without explicitly constructing $Q(\pmb{\xi})$~\cite{bravyi_mitigating_2021}.
The scalability of IBU for inferring observables in a similar manner remains questioned~\cite{bravyi_mitigating_2021}.
It is however easy to see that if there are no cross-talk in the measurement errors, and if the observables are dependent only on few bits, IBU can be applied locally.
In this case, we find that IBU mitigation~\eqref{eq:ibu} is comparable to the least square mitigation.

\section{Information theoretical consideration} 
We start with showing that the IBU update rule can be understood from the information theoretical viewpoint.
It will shortly become clear that this reformulation is actually very convenient for incorporation of prior structural information of the ideal outcome bitstrings into the inference.

Theoretically, one might question whether the sum of squares~\eqref{eq:sos} is an appropriate measure for the difference between probability distributions. 
For probability distributions, one might expect that the information theoretic relative entropy would be a more suitable measure~\cite{Cover1991a}. 
The relative entropy between two distributions $p(x)$ and $q(x)$ over the same sample space $\Omega$ is defined as $S[p(x)||q(x)] = \sum_{x \in \Omega} p (x) \ln p (x) / q(x)$. 
Although being asymmetric between $q(x)$ and $p(x)$, the relative entropy does resemble a metric between distributions in that it is positive and vanishes as $p(x)=q(x)$.
The relative entropy has been long known as an important operational quantity in information theory~\cite{Cover1991a}.
Its important role in statistical physics, and generally in physics, has been also appreciated~\cite{mackay_information_2003,mezard_information_2009}.

Replacing the sum of squared errors~\eqref{eq:sos} with the relative entropy, one obtains the least relative entropy estimate (LRE) for $Q(\pmb{\xi})$ as 
\begin{equation}
Q_{\RE} = \arg \min_{Q} S [P(\pmb{s}) \vert\vert \sum_{\pmb{\xi}} R(\pmb{s}|\pmb{\xi}) Q(\pmb{\xi}) ].
\label{eq:rev-ent}
\end{equation}
Generally, the probability distribution $Q(\pmb{\xi})$ can be characterised by a set of parameters $\theta$. 
Most often, the parameterisation of the distribution $Q_{\theta} (\pmb{\xi})$ imposes certain structural constraint on the distribution itself. 
In the extreme case, where there is no structure of $Q(\pmb{\xi})$ known, $\theta$ contains all the probability weights $Q(\pmb{\xi})$ themselves.
To indicate these parameters, the distribution will be subcripted by $\theta$ explicitly, thus denoted as $Q_{\theta} (\pmb{\xi})$. 
The optimisation~\eqref{eq:rev-ent} is therefore carried over to optimising these parameters $\theta$, 
$Q_{\RE} = Q_{\theta_{\RE}}$, where 
\begin{equation}
\theta_{\RE} =  \arg \min_{\theta} S [P(\pmb{s}) \vert\vert \sum_{\pmb{\xi}} R(\pmb{s}|\pmb{\xi}) Q_\theta (\pmb{\xi}) ].
\label{eq:rev-ent1}
\end{equation} 
Unlike~\eqref{eq:sos}, this is expressingly a nonquadratic optimisation problem.
Fortunately, optimisation problems of this type frequently appear in statistics and machine learning~\cite{mezard_information_2009,mackay_information_2003,Bishop2006a}, and there exists the so-called Expectation-Maximisation (EM) algorithm specifically devoted to them~\cite{Dempster1977a}.

\section{The EM algorithm}

\red{For the exposition of EM algorithm for the maximisation of the likelihood function, see~Ref.~\cite{Bishop2006a}, which has been also used to derive the IBU update rule~\eqref{eq:ibu} in~Ref.~\cite{srinivasan_scalable_2022}; see also~\cite{Lucy1974a,Volobouev2015a,Shepp1982a}. In this section, we are to explain the EM algorithm specifically adapted to the minimisation of the relative entropy~\eqref{eq:rev-ent}, or equivalently,~\eqref{eq:rev-ent1}. While they are closely related, this derivation highlights that the EM algorithm in this context is solely based on the decomposition of the conditional relative entropy.}

We start with recalling the chain rule for the relative entropy~\cite{Cover1991a}. For two given joint distributions $p(x,y)$ and $q(x,y)$ over $\Omega_X \times \Omega_Y$, the relative entropy can be decomposed as
\begin{align}
    S[p(x) || q(x)] = &  S[p(x,y)||q(x,y)]  \nonumber \\ 
                        & - \sum_{x \in \Omega_X} p (x) S [p (y|x) || q(y | x)].
                        \label{eq:chain}
\end{align}
The proof of this chain rule is a simple application of the chain rule for probability distributions, which can be found in Ref.~\cite{Cover1991a}. 

Coming back to the optimisation~\eqref{eq:rev-ent1}. The minimisation is complicated because of the marginal sum over the latent variable $\pmb{\xi}$. We would like then to rewrite this relative entropy between the marginal distributions over observed bitstrings $\pmb{s}$ in terms of the joint distributions over $\pmb{s}$ and $\pmb{\xi}$ using the chain rule~\eqref{eq:chain}. We are still left with the freedom to fix the joint distribution on the left variable of the relative entropy in~\eqref{eq:rev-ent1} to match the marginal $P(\pmb{s})$. 
Given an initial approximation $\theta^{(r)}$ for $\theta$, this joint distribution can be fixed in a natural way as follows.

With an initial approximation $\theta^{(r)}$ for $\theta$, we can define the joint distribution $Q_{\theta^{(r)}}(\pmb{s},\pmb{\xi}) = R (\pmb{s} | \pmb{\xi}) Q_{\theta^{(r)}} (\pmb{\xi})$, the marginal $Q_{\theta^{(r)}}(\pmb{s}) = \sum_{\pmb{\xi}} Q_{\theta^{(r)}}(\pmb{s},\pmb{\xi})$, and therefore the conditional distribution $Q_{\theta^{(r)}} (\pmb{\xi} | \pmb{s}) = Q_{\theta^{(r)}}(\pmb{s},\pmb{\xi}) / Q_{\theta^{(r)}} (\pmb{s})$. It is then natural to use the conditional distribution  $Q_{\theta^{(r)}} (\pmb{\xi} | \pmb{s}) $ to define the joint distribution $P_{\theta^{(r)}} (\pmb{s},\pmb{\xi}) =  P(\pmb{s}) Q_{\theta^{(r)}} (\pmb{\xi} | \pmb{s}) $.
With this joint distribution $P_{\theta^{(r)}} (\pmb{s},\pmb{\xi})$, we apply the chain rule~\eqref{eq:chain} and find
\begin{align}
    S[P(\pmb{s}) || Q_{\theta} (\pmb{s})] = & S [P_{\theta^{(r)}} (\pmb{s},\pmb{\xi}) || Q_{\theta} (\pmb{s},\pmb{\xi})]  \nonumber \\ 
    & -\sum_{\pmb{s}} P(\pmb{s}) S[P_{\theta^{(r)}}( \pmb{\xi} | \pmb{s}) || Q_{\theta}( \pmb{\xi} | \pmb{s})],
    \label{eq:joint}
\end{align}
where $Q_{\theta} (\pmb{s},\pmb{\xi}) = R(\pmb{s}|\pmb{\xi}) Q_\theta (\pmb{\xi})$ and $Q_{\theta} (\pmb{s}) = \sum_{\pmb{\xi}} Q_{\theta} (\pmb{s},\pmb{\xi})$. We see that in passing from the relative entropy between the marginal distributions~\eqref{eq:rev-ent1} to that between the joint distributions, an additional term arises in~\eqref{eq:joint}. However, notice that $P_{\theta^{(r)}}( \pmb{\xi} | \pmb{s}) = Q_{\theta^{(r)}}( \pmb{\xi} | \pmb{s})$ by construction, and thus this term vanishes for $\theta=\theta^{(r)}$. Therefore, if we opt to minimise the relative entropy between the joint distributions as the next approximation for $\theta$, 
\begin{equation}
    \theta^{(r+1)} = \arg \min_{\theta} S [P_{\theta^{(r)}} (\pmb{s},\pmb{\xi}) || Q_{\theta} (\pmb{s},\pmb{\xi})], 
    \label{eq:em-update}
\end{equation}
this additional term in~\eqref{eq:joint} contributes a negative value to $S[P(\pmb{s}) || Q_{\theta^{(r+1)}} (\pmb{s})]$. As a result, the relative entropy between the marginals decreases, 
\begin{equation}
S[P(\pmb{s}) || Q_{\theta^{(r+1)}} (\pmb{s})] \le S[P(\pmb{s}) || Q_{\theta^{(r)}} (\pmb{s})] .
\end{equation}

To summarise, the EM algorithm then substitutes the minimisation~\eqref{eq:rev-ent1} by ~\eqref{eq:em-update}. 
While there is difference between ~\eqref{eq:rev-ent1} and~\eqref{eq:em-update}, the relative entropy ~\eqref{eq:rev-ent1} decreases in every step. 
Therefore the iterative procedure~\eqref{eq:em-update} converges to a local minimum of~\eqref{eq:rev-ent1}.
Randomisation of the initial starting point can be used to asset the globality of the obtained minimum. 
Notice that in the minimisation~\eqref{eq:em-update}, the summation over the latent variable $\pmb{\xi}$ does not present. 
As a result, it is often a straightforward parameter estimation problem in statistics.  

Using the explicit formula for the distributions in~\eqref{eq:em-update}, one finds 
\begin{equation}
    \theta^{(r+1)} = \arg \min_{\theta} \sum_{\pmb{s}} P(\pmb{s}) \sum_{\pmb{\xi}} Q_{\theta^{(r)}}(\pmb{\xi} | \pmb{s}) \ln  R(\pmb{s} \vert \pmb{\xi}) Q_{\theta}(\pmb{\xi})
    \label{eq:EM}
\end{equation}
where $Q_{\theta^{(r)}}(\pmb{\xi} | \pmb{s}) = R(\pmb{s}|\pmb{\xi}) Q_{\theta^{(r)}}(\pmb{\xi})/[\sum_{\pmb{\xi}'} R(\pmb{s}|\pmb{\xi}') Q_{\theta^{(r)}}(\pmb{\xi}')]$.
Taking the case where $\theta$ simply contains all the probability weights of $Q(\pmb{\xi})$, the update rule~\eqref{eq:EM} leads directly to
\begin{equation}
    Q^{(r+1)} (\pmb{\xi}) = \sum_{\pmb{s}} P(\pmb{s}) \frac{ R(\pmb{s}|\pmb{\xi}) Q^{(r)}(\pmb{\xi})}{\sum_{\pmb{\xi}'} R(\pmb{s}|\pmb{\xi}') Q^{(r)}(\pmb{\xi}')}.
\end{equation}
This is precisely the IBU update rule~\eqref{eq:ibu}.
Using the finite sampling approximation~\eqref{eq:statistics} for $P(\pmb{s})$, one has
\begin{equation}
    Q^{(r+1)} (\pmb{\xi}) = \frac{1}{M}\sum_{\mu=1}^{M}  \frac{ R(\pmb{s}_\mu|\pmb{\xi}) Q^{(r)}(\pmb{\xi})}{\sum_{\pmb{\xi}'} P(\pmb{s}_\mu|\pmb{\xi}') Q^{(r)}(\pmb{\xi}')}.
\end{equation}

\section{Structural mitigation and unstructural mitigation of measurement readout errors}

\subsection{Structural mitigation of distributions}
With the information theoretic basis~\eqref{eq:rev-ent1} of the IBU update rule~\eqref{eq:ibu}, it is clear how to incorporate the structural information of the ideal distribution $Q(\pmb{\xi})$ into its inference.
Suppose $Q(\pmb{\xi})$ is supported on at most $K$ different (unknown) bitstrings $\{\pmb{\xi}_\nu\}_{\nu=1}^{K}$ with corresponding probability weights $\{p_{\nu}\}_{\nu=1}^{K}$. This means that $Q(\pmb{\xi})$ can be written as 
\begin{equation}
Q(\pmb{\xi}) = \sum_{\nu=1}^{K} p_{\nu} \delta_{\pmb{\xi} ,\pmb{\xi}_\nu}.   
\label{eq:structural-dist}
\end{equation}
Here $K$ is of order $\mathcal{O}(1)$, and generally $K \ll M \ll 2^n$.
The maximisation~\eqref{eq:EM} leads to
\begin{equation}
    \{p_\nu,\pmb{\xi}_\nu\}_{\RE}= \min_{\{p_\nu,\pmb{\xi}_\nu\}} \sum_{\mu=1}^{M} \ln [\sum_{\nu} P(\pmb{s}_{\mu}|\pmb{\xi}_\nu) p_\nu].
    \label{eq:re1}
\end{equation}
It fact, the model~\eqref{eq:structural-dist} belongs to a general class known as mixture models in machine learning~\cite{Bishop2006a}.  
In this case, direct application of the EM algorithm~\eqref{eq:EM} to the minimisation~\eqref{eq:re1} leads to the iterative update rule
\begin{align}
p^{(r+1)}_{\nu} &=   \frac{1}{M}\sum_{\mu} \gamma^{(r)} (\pmb{\xi}_\nu \vert \pmb{s}_{\mu}) \\
\pmb{\xi}_\nu^{(r+1)} & = \arg \max_{\pmb{\xi}_\nu} \sum_{\mu} \gamma^{(r)} (\pmb{\xi}_\nu \vert \pmb{s}_{\mu}) \ln R(\pmb{s}_\mu \vert \pmb{\xi}_\nu),
\label{eq:mixture-model}
\end{align}
where $\gamma^{(r)}(\pmb{\xi}_\nu \vert \pmb{s}_{\mu}) =  R(\pmb{s}_\mu \vert \pmb{\xi}_\nu) p^{(r)}_{\nu} / [\sum_{\nu'=1}^{K} R(\pmb{s}_{\mu} \vert \pmb{\xi}_{\nu'}) p^{(r)}_{\nu'}]$. 
Like the iterative update rule of the IBU~\eqref{eq:ibu}, in every step the relative entropy~\eqref{eq:rev-ent} decreases.

\begin{figure}
\includegraphics[width=0.45\textwidth]{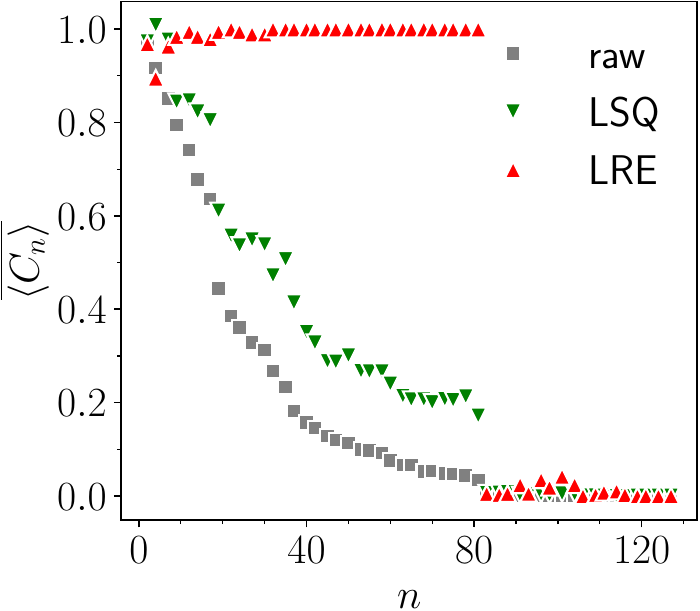}
\caption{Structural mitigation inferring the hidden bitstrings based on measured bitstrings of the GHZ state prepared in a quantum computer based on the least relative entropy (LRE). The global correlation $\mean{C_n} = \mean{\prod_{k=1}^{n} \xi_k}+ \mean{\prod_{k=1}^{n} \bar{\xi}_k}$ is averaged over $190$ bootstraps out of the $100000$ original shots, each with $2000$ shots to obtained $\overline{\mean{C_n}}$. For comparison, unstructual mitigation using the least square scheme (LSQ) and the raw data are also presented.}
\label{fig:structural}
\end{figure}

{\it  Illustration of structural mitigation using the GHZ dataset.}
We use the data of the simulation of the GHZ state on the 127-qubit quantum computer Washington at IBM published in Ref.~\cite{srinivasan_scalable_2022} to mimic the structural mitigation problem. Imagine that this is an output of a nearly deterministic quantum computation, which outputs $K=2$ possible outcome bitstrings. 
For the exact GHZ state, the two ideal bitstrings are $(000\ldots 0)$ and $(111\ldots 1)$. 
Readout noises would alter the bits with certain rates, which are measured in the calibration process.
Without assuming the GHZ state, the algorithm~\eqref{eq:mixture-model} allows us to reconstruct the two hidden bitstrings from the noisy observed bitstrings. Figure~\ref{fig:structural} presents the obtained total probability weight of $(000\ldots 0)$ and $(111 \ldots 1)$, which is given by $\mean{C_n} = \mean{\prod_{k=1}^{n} \xi_k}+ \mean{\prod_{k=1}^{n} \bar{\xi}_k}$, where $\bar{\xi}_k$ denotes the flipped bit value of $\xi_k$. 
One observes that for the experiments with number of qubits smaller than $80$, nearly perfect reconstruction of the hidden bitstrings are obtained. 
One may also assume that the number $K$ of hidden bitstrings is unknown and vary $K$ accordingly.
We increases $K$ upto $10$ and found that the results remain essentially unchanged.
Moreover, for each experiment we resample only $M=2000$ shots from the $100000$ available original shots, illustrating that inferring structural distribution is an easy classical error correction.
The computation can also be carried out in a normal laptop. 
For the reference, we plot the same quantity using the unstructural mitigation algorithm~\eqref{eq:sos}, which produces a much lower success probability.
As we mentioned, however, this should not be considered as a fair comparison, since unstructural inference addresses a different problem. 

For experiments with the numbers of qubits larger than $80$, the reconstruction of the hidden bitstrings suddenly fails. 
It has been also noticed in the original Ref.~\cite{srinivasan_scalable_2022} that other mitigation methods altogether fail at this point. 
This is perhaps due to the high accumulation of the gate errors and the output quantum states are actually far from a GHZ state.

\subsection{Unstructural mitigation of local observables}

As we argued, for the unstructural mitigation, the principal interest shifts from inferring the probability distribution $Q(\pmb{\xi})$ to inferring certain observables. 
Also, most often, not any observable is of interest.
The most interesting observables are often decomposed into a linear combination of simpler observables, which are local in the sense that they are functions of only few outcome bits in the bitstring.
In this case, it is easily seen that under the independent noise model~\eqref{eq:noise}, the reconstruction~\eqref{eq:sos} and~\eqref{eq:rev-ent} can be restricted locally to a subset of bits, independent of the others.
It means that the same algorithms can be used to infer any marginal distributions over smaller subsets of bits, with which the observables can be computed.  
Notice that this is the non-trivial feature of the independent noise model.

\begin{figure}
\includegraphics[width=0.45\textwidth]{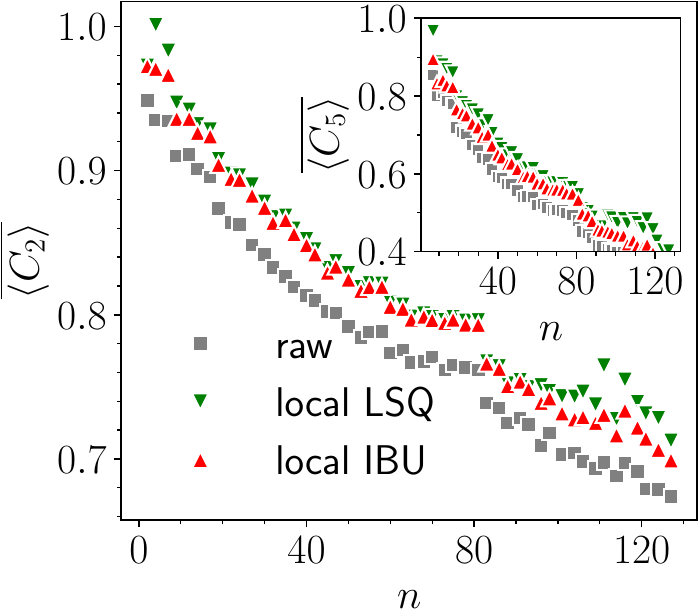}
\caption{
Unstructural mitigation of local observables based on measured bitstrings of the GHZ state prepared in a quantum computer. 
A fixed number $300$ subgroups of bits of size $l$ (with $l=2$ and $l=5$), $(i_1,i_2,\ldots,i_l)$,  are randomly chosen. 
For each sampled subgroup of bits, the marginal distributions are obtained by least sum of squared errors (LSQ) and Iterative Bayesian Unfolding (IBU). 
The local correlation $C_{l} (i_1,i_2,\ldots,i_l)$ is computed and $\overline{\mean{C_l}}$ is obtained by averaging over the $300$ sampled bit groups. 
}
\label{fig:unstructural}
\end{figure}

{\it  Unstructural learning using the GHZ dataset.}
To illustrate how this works, we use the same the data of the quantum simulation of the GHZ state as described in the previous section. 
Instead of inferring the distribution over the hidden outcome bitstrings $Q(\pmb{\xi})$, we infer marginal distributions over group of bits of length $l$ using either~\eqref{eq:sos} or~\eqref{eq:rev-ent}. 
The inferred marginal distribution are used to compute the local correlations $\mean{C_l(i_1,i_2,\ldots,i_l)}= \mean{\prod_{k=1}^{l} \xi_k} + \mean{\prod_{k=1}^{l} \bar{\xi}_k}$. We use $\overline{\mean{C_l}}$ to denote its average over random samples of groups of $l$ bits in the bitstring. 
The local correlations are presented in Figure~\ref{fig:unstructural}. 
One sees that the local correlations are well reconstructed for all experiments, with only a minor difference between local least square error mitigation and least relative entropy mitigation.

\section{Discussion and conclusion}

We have shown that the IBU algorithm can be understood from an information theoretical viewpoint.
\red{While conceptually simple, we hope that this can also brings a different perspective on the theoretical analysis of IBU for mitigation of experimental errors beyond quantum computer readouts.}
\red{The derivation in fact shows that structural information can be easily integrated to the mitigation of readout errors in quantum computers in a much natural way, which is so far unknown for LSQ.}
For unstructural mitigation, we show that as long as local observables are concerned, both IBU and LSQ perform nearly equally.  
Our present analysis assumes no cross-talk in the readout errors of different qubits. 
As it is pointed out recently~\cite{bravyi_mitigating_2021}, cross-talks might play an important role in certain quantum computers. 
It is an interesting future project to extend our analysis to this case. 

\begin{acknowledgments}
I would like to thank  
Johannes Berg,
Claus Grupen,
and Otfried G\"uhne
for discussions and comments.
Matthias Kleinmann's many questions are highly appreciated.
My thanks to the authors of Ref.~\cite{srinivasan_scalable_2022} for providing me with their data, in particular, Srinivasan for his kind instruction.
I am grateful to Qiongyi He for hosting me at the University of Beijing, during which this manuscript was completed.
The University of Siegen is acknowledged for enabling our computation through the \texttt{OMNI} cluster.
This work was supported by 
the Deutsche Forschungsgemeinschaft (DFG, German Research Foundation, project numbers 447948357 and 440958198), 
the Sino-German Center for Research Promotion (Project M-0294), 
the German Ministry of Education and Research (Project QuKuK, BMBF Grant No. 16KIS1618K) 
and the ERC (Consolidator Grant 683107/TempoQ).
\end{acknowledgments}

\bibliography{refs.bib}

\end{document}